\documentclass[a4paper,floatfix,twocolumn,prc,nofootinbib]{revtex4}
\usepackage{url}
\usepackage{graphicx}
\usepackage{epsfig}
\usepackage{gensymb}
\usepackage{amsmath}
\usepackage{amsfonts}
\usepackage{amssymb}
\usepackage{mathrsfs}
\usepackage{xcolor}
\usepackage{natbib}
\usepackage[normalem]{ulem}

\newcommand{\F}{\text{\tiny{F}}}
\newcommand{\Fc}{\text{\tiny{F,c}}}
\newcommand{\Lif}{\text{\tiny{L}}}
\newcommand{\Lut}{\text{\tiny{L}}}
\newcommand{\M}{\text{\tiny{M}}}
\newcommand{\FFLO}{\text{\tiny{FFLO}}}

\DeclareMathOperator{\Real}{\mathcal{R}e}
\DeclareMathOperator{\Imag}{\mathcal{I}m}

\newcommand{\Ref}[1]{Ref.~\cite{#1}}

\newcommand{\Sec}[1]{Sec.~\ref{#1}}

\usepackage{etoolbox}
\newbool{hardcopybool}

  \boolfalse{hardcopybool}

\ifdefined\hardcopy 
  \booltrue{hardcopybool}
\fi
\ifbool{hardcopybool}{\typeout{!!! PRINTING HARD COPY}}{}

\usepackage[colorlinks=true,
  \ifbool{hardcopybool}%
         {linkcolor=black,citecolor=black,filecolor=black,urlcolor=black}%
         {linkcolor=black,citecolor=black,filecolor=black,urlcolor=black}]{hyperref}

\hypersetup{colorlinks=true,linkcolor=magenta, citecolor=blue}

\DeclareMathOperator{\e}{e}

\DeclareMathOperator{\ii}{i}

\DeclareMathOperator{\kk}{\textbf{\text{k}}}
\DeclareMathOperator{\pp}{\mathbf{p}}

\begin{document}

\title{Application of renormalized RPA to polarized Fermi gases}

\author{David Durel}
\email{dureldavid@ipno.in2p3.fr}

\author{Michael Urban}
\email{urban@ipno.in2p3.fr}

\affiliation{Institut de Physique Nucl\'{e}aire, CNRS/IN2P3, Univ. Paris-Sud 11, Universit\'{e} Paris-Saclay, F-91406 Orsay Cedex, France}

\begin{abstract}
\noindent
We consider a spin imbalanced Fermi gas at zero temperature in the
normal phase on the BCS side of the BCS-BEC crossover and
  around unitarity. We compute the critical polarization for pairing,
the correlated occupation numbers and the contact in an
  extension of particle-particle RPA (also called
  non self-consistent \textit{T}-matrix approach or ladder
  approximation). The so-called renormalized RPA consists in
computing the \textit{T} matrix with self-consistently determined
occupation numbers. The occupation numbers are determined
  either by keeping the self-energy only to first order or by
  resumming the Dyson equation. In this way, the result for the
critical polarization, strongly overestimated in standard RPA, is
clearly improved. We also discuss some problems of this approach.
\end{abstract}

\maketitle

\section{Introduction}
Cold atoms have allowed for the first realization of the crossover
from superfluidity of Cooper pairs as described by the
Bardeen-Cooper-Schrieffer (BCS) theory to Bose-Einstein condensation
(BEC) of dimers. This is possible because the scattering length $ a $,
i.e., the interaction strength, can be changed thanks to the Feshbach
resonance. At zero temperature, a Fermi gas with two spin states $ \sigma = 
\uparrow, \downarrow $ of equal masses and populations is
superfluid at any interaction strength. This is not the case for a
polarized gas, in which pairing disappears beyond some critical
polarization $ P_{c} $ whose value depends on the interaction
strength. In the polarized case, other forms of pairing may exist such
as the Fulde-Ferrell-Larkin-Ovchinnikov (FFLO) phase~\cite{Fulde1964,
  Larkin1964} characterized by an oscillating order parameter
corresponding to pairs with a non-zero total momentum. So
  far, in experiments on polarized Fermi gases performed on resonance
  ($ a \to \infty $, the so-called unitary limit)~\cite{Partridge2006,
    Shin2006, Shin2008} and in the BCS-BEC
  crossover~\cite{Shin2008prl, Olsen2015}, the FFLO phase was not
  seen, but a phase separation into a paired and an unpaired state was
  found. In fact, the FFLO phase may be difficult to see in a harmonic
  trap and the use of flat traps may clarify this question in the
  future. The problem of pairing in asymmetric systems has also been
studied, e.g., in the case of proton-neutron pairing in asymmetric
nuclear matter~\cite{Stein2014} and in the context of QCD with color
superconductivity in the case of quark matter which also involves
particles of different masses~\cite{Alford2008}.

Many theoretical studies have already addressed the problem of pairing
in polarized Fermi gases, for reviews see~\cite{Radzihovsky2010,
  Chevy2010, Gubbels2013}. One problem of polarized systems is that
the standard Nozi\`{e}res and Schmitt-Rink (NSR)~\cite{Nozieres1985}
approach, which describes the BCS-BEC crossover at finite temperature,
fails in the polarized case~\cite{Liu2006, Parish2007}. Different
variants of the NSR approach were proposed to solve this
problem~\cite{Kashimura2012, Chen2007, He2007, Pantel2014}. All these
approaches have in common to be based on the ladder approximation for
the in-medium \textit{T} matrix and to include the self-energy in a
somewhat more self-consistent way than it is done in the NSR
theory.

In this paper, we consider only the case of zero
  temperature and the normal phase. This implies that the polarization
  of the gas must exceed the critical polarization $P_c$ below which
  superfluidity sets in. The ladder approximation becomes then
equivalent to what is known in nuclear physics as the
particle-particle Random-Phase Approximation (pp-RPA) which was
applied to polarized Fermi gases in~\cite{Urban2014}. The pp-RPA gives
satisfactory results on the BCS side for not too strong interaction,
but at unitarity it overestimates strongly the critical
polarization. Within this formalism, the onset of
  superfluidity appears as an instability, but it is not possible to
  describe the superfluid phase. Maybe one could generalize the
  formalism to the superfluid phase using Gor'kov Green's functions,
  similarly to what was done in the unpolarized case at finite
  temperature, e.g., in~\cite{Pieri2004}, but this is beyond the scope
  of the present study.

From a general perspective, the RPA describes correlations in the
medium, e.g., the correlation energy or correlated occupation numbers,
starting from an uncorrelated ground state. In this sense, the RPA is
not fully consistent. Different extensions of RPA were developed to
take into account these correlations in a more consistent way. One of
them is the self-consistent RPA (SCRPA)~\cite{Dukelsky1998,
  Schafer1999, Hirsch2002}, which is based on the correlated RPA
ground state. In practice, however, the SCRPA is very difficult to
implement except in simple toy models. An approximation to the SCRPA
is the renormalized RPA (r-RPA)~\cite{Catara1996, Delion2005}, which
instead of using the correlated ground state uses only the correlated
occupation numbers in the calculation. The inclusion of the correlated
occupation numbers can be expected to solve, at least partially, the
problem of the critical polarization in the polarized Fermi gas at
zero temperature.

Our article is organized as follows. In \Sec{sec:self-consistent}, we recall briefly the pp-RPA formalism and describe the basic idea of the self-consistent processing of the occupation numbers. We use two different methods to obtain the occupation numbers, one based on the Dyson equation truncated at first order that we call RPA(1st), and the other using the full Dyson equation that we call RPA($ \infty $). We will discuss the results obtained for the occupation numbers and the contact. In  \Sec{sec:Critical_polarization}, we present the calculation of the critical polarization, first in the case of RPA and then in the case of the r-RPA. Finally, we conclude in \Sec{sec:conclusions}.

\section{Self-consistent occupation numbers}
\label{sec:self-consistent}
\subsection{Recapitulation of pp-RPA at zero temperature}
\label{sec:Recall_of_pp-RPA}
Let's start by recalling the pp-RPA for a zero-temperature polarized
fermion gas. At zero temperature, it is common to fix the densities of up and down populations, denoted $
\rho_{\uparrow} $ and $ \rho_{\downarrow} $, unlike at finite
temperature where one usually fixes the chemical
  potentials. The polarization $ P $ of the gas is then defined as
\begin{equation}\label{polarisation}
P = \dfrac{\rho_{\uparrow} - \rho_{\downarrow}}{\rho_{\uparrow} + \rho_{\downarrow}} \, ,
\end{equation}
assuming $ \rho_{\uparrow} > \rho_{\downarrow} $ as
convention. Since the formalism breaks down at the
  superfluid phase transition (see
  Sec.~\ref{sec:Critical_polarization}), we limit ourselves to
  sufficiently strong polarizations where the system remains normal
  fluid even at zero temperature.

\begin{figure}
\includegraphics[scale=0.31]{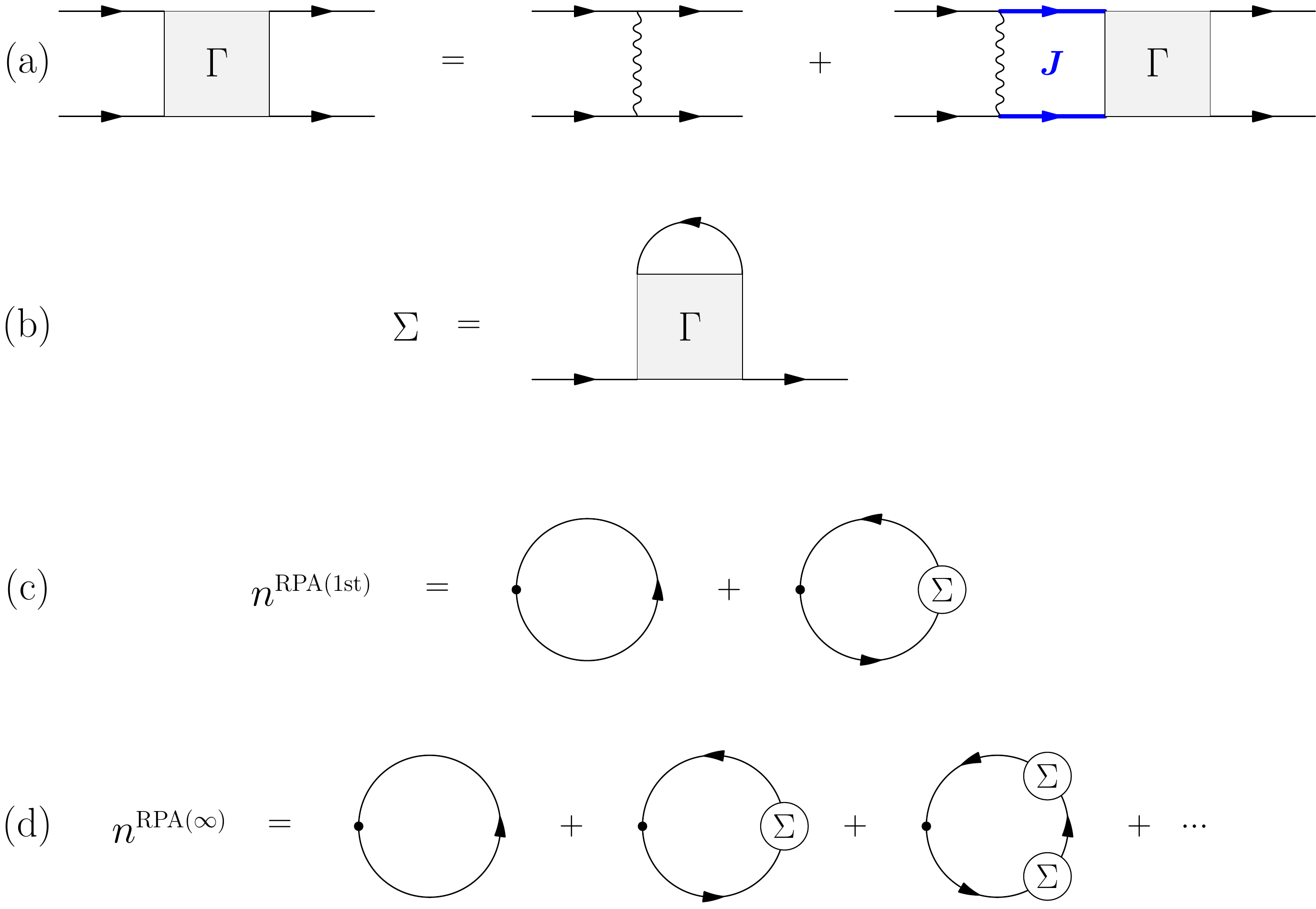}
\caption{Representation in terms of Feynman diagrams of (a)
  the vertex function $ \Gamma $, (b) the self-energy $ \Sigma $, and
  the occupation numbers $ n^{\sigma} $ calculated with (c) the
  truncated Dyson equation [RPA(1st)] and (d) the full Dyson equation
  [RPA($ \infty $)]. Inserting self-consistent occupation numbers into
  the two-particle propagator $J$ within the r-RPA corresponds roughly
  to an approximate way of dressing the thick blue lines in diagram (a), but
  there is no one-to-one correspondence in terms of diagrams.}
\label{graphe_feynman}
\end{figure}

The pp-RPA is based on the formalism of the in-medium \textit{T} matrix which is written as the sum of ladder diagrams that we can see in Fig.~\ref{graphe_feynman}~(a).

In the case of cold atoms, it is appropriate to take a contact interaction with coupling constant $ g $. The vertex function $ \Gamma $ is written as
\begin{equation}
\Gamma (\kk , \, \omega) = \dfrac{1}{1/g - J (\kk , \, \omega)} \, ,
\end{equation}
with $ J (\kk , \, \omega) = J_{hh} (\kk , \, \omega) + J_{pp} (\kk , \, \omega) $, including both particle-particle (pp) and hole-hole (hh) propagation.
The function $ J $ needs to be regularized~\cite{Perali2002} which gives the following expressions:
\begin{gather}
  g = \dfrac{4 \pi a}{m} \, ,\\
  J_{hh}(\kk, \omega) = - \int \dfrac{\mathrm{d}^{3}\mathbf{p}}{(2 \pi)^{3}} \, \dfrac{n^{\uparrow}(| \frac{\kk}{2}+\pp |) \, n^{\downarrow}( | \frac{\kk}{2}-\pp |) }{\omega - \varepsilon_{\frac{\kk}{2}+\pp} - \varepsilon_{\frac{\kk}{2}-\pp} - \ii \eta} \, ,\label{J_hh_auto}\\
J_{pp}(\kk, \omega) = \int \dfrac{\mathrm{d}^{3}\mathbf{p}}{(2 \pi)^{3}} \left( \dfrac{\bar{n}^{\uparrow}( | \frac{\kk}{2}+ \pp | ) \, \bar{n}^{\downarrow}( | \frac{\kk}{2}-\pp |) }{\omega - \varepsilon_{\frac{\kk}{2}+\pp} - \varepsilon_{\frac{\kk}{2}-\pp} + \ii \eta} + \dfrac{m}{p^{2}} \right) \, ,\label{J_pp_auto}
\end{gather}
where $ \varepsilon_{\kk} = k^{2}/(2m) $, $
  n^{\sigma} $ are the occupation numbers of spin $\sigma$, and $
  \bar{n}^{\sigma} = 1-n^{\sigma} $. In the case of standard RPA, the
  expression of $ n^{\sigma} $ is $ n^{\sigma}(k) = \theta
  (k_{\F}^{\sigma} - k) $.

To calculate the correlated occupation numbers $ n^{\sigma} $ in
standard RPA, one uses the Dyson equation truncated at first order
[Fig.~\ref{graphe_feynman}~(c)], i.e.,
\begin{equation}\label{fonction_green_ordre1}
G^{\sigma} = G_{0}^{\sigma} + G_{0}^{\sigma} \Sigma^{\sigma} G_{0}^{\sigma} \, ,
\end{equation}
where $ G^{\sigma} $ is the dressed Green's function, $ G_{0}^{\sigma} $ the bare Green's function and $ \Sigma^{\sigma} $ the self-energy represented in Fig.~\ref{graphe_feynman}~(b), which has the form
\begin{equation}
\Sigma^{\sigma}(\kk, \, \omega) = - \ii \int\! \dfrac{\mathrm{d}^{3} \mathbf{p}}{(2 \pi)^{3}} \int\! \dfrac{\mathrm{d} \omega'}{2\pi} G_{0}^{\bar{\sigma}}(\pp, \, \omega') \, \Gamma(\kk + \pp , \, \omega + \omega') \, ,
  \label{eq:Sigma_general}
\end{equation}
where $ \bar{\sigma} $ denotes the spin opposite to $ \sigma $.
This is the same approximation that is used at finite temperature in the NSR theory~\cite{Nozieres1985}. The general expression for the occupation numbers reads~\cite{Fetter1971}
\begin{equation}
n^{\sigma} (k) = - \ii \int \dfrac{\mathrm{d} \omega}{2 \pi} \, \e^{\ii \omega \eta} G^{\sigma}(\kk, \omega) \, .
\end{equation}
The resulting expressions for the occupation numbers of the holes and particles are respectively~\cite{Urban2014}
\begin{multline}\label{n_hole}
n^{\sigma}(k < k_{\F}^{\sigma}) = 1 + \int \! \dfrac{\mathrm{d}^{3} \mathbf{p}}{(2 \pi)^{3}} \int_{\Omega_{\F}}^{+ \infty} \dfrac{\mathrm{d} \omega}{\pi} \, \dfrac{\theta(k_{\F}^{\bar{\sigma}} - |\pp - \kk|)}{(\omega - \varepsilon_{\kk} - \varepsilon_{\pp-\kk})^{2}} \\
 \times \Imag \Gamma(\pp, \, \omega) \, ,
\end{multline}
\begin{multline}\label{n_particle}
n^{\sigma}(k > k_{\F}^{\sigma}) = - \int \dfrac{\mathrm{d}^{3} \mathbf{p}}{(2 \pi)^{3}} \, \int_{- \infty}^{\Omega_{\F}} \dfrac{\mathrm{d} \omega}{\pi} \, \dfrac{\theta(|\pp - \kk| - k_{\F}^{\bar{\sigma}})}{(\omega - \varepsilon_{\kk} - \varepsilon_{\mathbf{p}-\kk})^{2}} \\
 \times \Imag \Gamma(\pp, \, \omega) \, ,
\end{multline}
where $ \Omega_{\F} = (k_{\F}^{\uparrow \, 2} + k_{\F}^{\downarrow \, 2})/(2m) $ is the energy separating the two-particle and two-hole continua.
The occupation numbers that we obtain can be decomposed into a continuous part $ n_{c} $ and a step of height $ Z^{\sigma} = \lim\limits_{\varepsilon \rightarrow 0} n^{\sigma}(k_{\F}^{\sigma}-\varepsilon) - n^{\sigma}(k_{\F}^{\sigma}+\varepsilon) $, i.e.,
\begin{equation}\label{n_auto}
n^{\sigma}(k) = Z^{\sigma} \, \theta(k_{\F}^{\sigma}-k) + n_{c}^{\sigma}(k) \, .
\end{equation}
On the one hand, the procedure of keeping in Eq.~\eqref{fonction_green_ordre1} only the first order in the self-energy makes it possible to satisfy the Luttinger theorem, i.e., that the correlations do not modify the densities of the gas. On the other hand, it leads to a pathology when the interaction becomes too strong, the height of the step $ Z $ becoming negative (and occupation numbers can become negative, too)~\cite{Urban2014}.

\subsection{Renormalized pp-RPA}
\label{sec:Renormalized_pp-RPA}
The idea of renormalized RPA (r-RPA) is to reuse the pp-RPA formalism presented in the preceding subsection but in which the occupation numbers in Eqs.~\eqref{J_hh_auto} and~\eqref{J_pp_auto} are no longer Heaviside functions but the correlated occupation numbers writted in Eq.~\eqref{n_auto}. This approximation can be justified using the equation-of-motion method for the two-particle Green's function, see, e.g., \cite{Dukelsky1998}.

In some sense, this prescription can be viewed as an approximation to
the two-particle Green's function $ J $ one would obtain by dressing
the single-particle propagators appearing in the ladder diagrams in
Fig.~\ref{graphe_feynman}~(a). However, there are important
differences. For instance, in the limit $ P \to 1 $, the r-RPA reduces
to the RPA because the up and down occupation numbers tend to
$\theta(k_{\F}^{\uparrow} - k)$ and $ 0 $, respectively, while the
dressed propagator of the down particle (polaron) would remain
non-trivial even in this limit.

\begin{figure}[!h]
\includegraphics[scale=1.0]{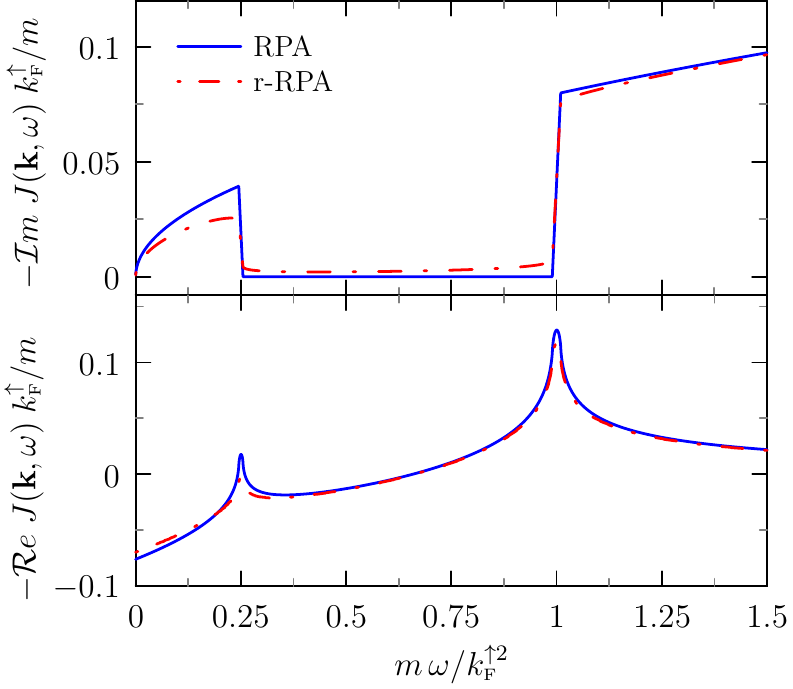}
\caption{Imaginary and real part of $ J $ as function of $ \omega $ for $ k = 0.01 \, k_{\F}^{\uparrow} $ and $ k_{\F}^{\downarrow} = k_{\F}^{\uparrow}/2 $. The blue solid lines represent $ J $ with uncorrelated occupation numbers and the red dash-dotted lines are obtained with self-consistent occupation numbers for $ a \, k_{\F}^{\uparrow} = -2 $.}
\label{representation_J}
\end{figure}

By including the correlations ($ Z^{\sigma} < 1 $) in the calculation of the two-particle propagator $ J $, the logarithmic singularity of $ \Real J $, responsible for the instability of the normal phase~\cite{Fetter1971}, will be reduced. This can be seen in Fig.~\ref{representation_J}, where we show $ J(\kk, \omega) $ for small non-vanishing $ \kk $ for better visibility.
This softening of the singularity will allow the normal phase to remain stable at lower polarization.

\begin{figure}[!h]
\includegraphics[scale=1.0]{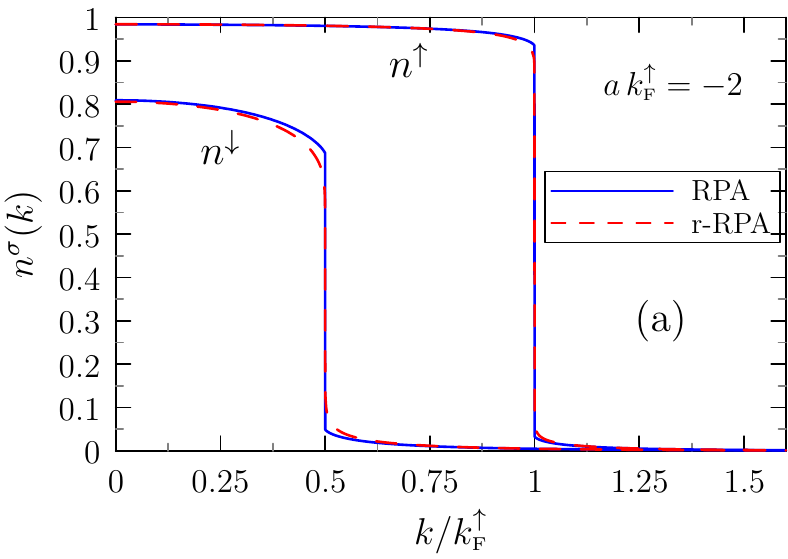}
\includegraphics[scale=1.0]{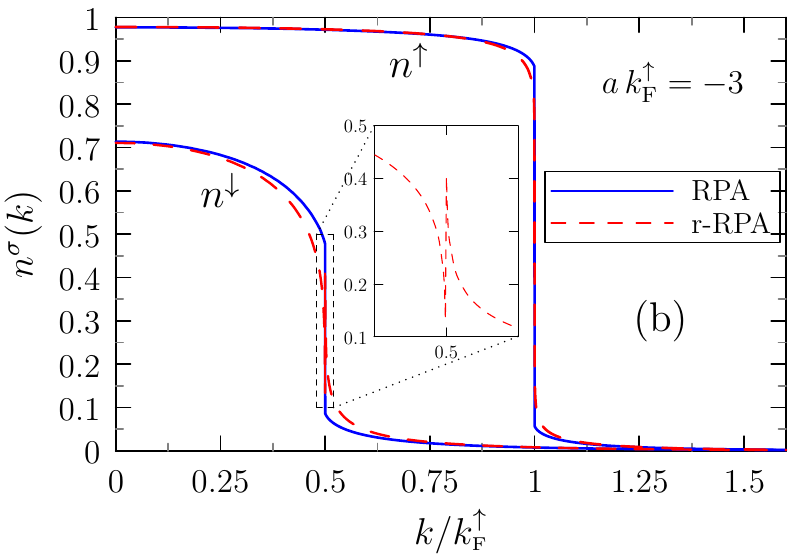}
\caption{Up and down occupation numbers for the polarization $
  k_{\F}^{\downarrow} = k_{\F}^{\uparrow}/2 $ ($ P = 0.778 $) for the
  RPA (blue solid line) and the r-RPA (at convergence, red dashed
  line). (a) $ a \, k_{\F}^{\uparrow} = -2 $, (b) $ a \,
  k_{\F}^{\uparrow} = -3 $. The inset shows a zoom on the
    unphysical negative step ($Z^\downarrow<0$) of the down occupation
    numbers.}
\label{nombre_occupation_r-rpa}
\end{figure}

With this formalism, the results of the pp-RPA can be considered as the first iteration of the self-consistent calculation. To carry out the iteration, the correlated occupation numbers are calculated according to Eqs.~\eqref{n_hole} and~\eqref{n_particle} and are reinjected into the functions $ J $, Eqs.~\eqref{J_hh_auto} and~\eqref{J_pp_auto}. This procedure is repeated until convergence is reached. One important thing to notice is that the converged result is independent of the initial occupation numbers. The comparison between the RPA and r-RPA occupation numbers is shown in Fig.~\ref{nombre_occupation_r-rpa} for different values of the interaction strength.
For weak interactions, i.e., $ | a \, k_{\F}^{\uparrow} | < 1 $, the r-RPA provides virtually no correction to the RPA. In contrast, for stronger interactions, ​​the r-RPA reduces the value of $ Z $ more strongly than the RPA calculation. Therefore the self-consistent treatment does not cure the pathology of RPA that the value of $ Z $ becomes negative for interactions that are too strong; on the contrary, the negative step appears already for weaker interactions (for instance, in Fig.~\ref{nombre_occupation_r-rpa}(b), the step of $ n^{\downarrow}(k) $ is negative in r-RPA as can be seen in the zoom).

\begin{figure}[!h]
\includegraphics[scale=0.8]{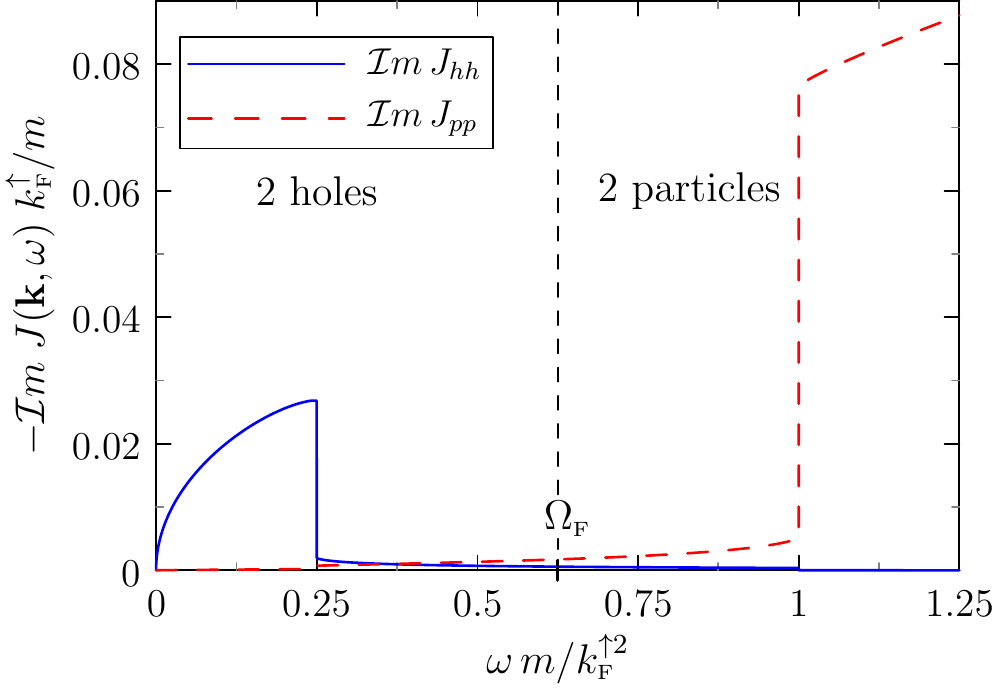}
\caption{Imaginary part of $ J $ as function of $ \omega $ for $ \kk = \boldsymbol{0} $ and $ k_{\F}^{\downarrow} = k_{\F}^{\uparrow}/2 $. The blue line is the imaginary part of $ J $ for holes and the red dashed line for particles. As can be seen, each curve extends into the energy zone of the other.}
\label{J_autocoherent}
\end{figure}

Let us note that it follows from the spectral representation of the two-particle Green function $ J $ (see chapter 15.2 of~\cite{Blaizot1986}) that in principle the two-hole continuum of $ \Imag J_{hh} $ should be restricted to energies below $ \Omega_{\F} $ and the two-particle continuum of $ \Imag J_{pp} $ to energies above $ \Omega_{\F} $, as it is the case in standard pp-RPA. In the self-consistent treatment, i.e., by including the correlations in the calculation of $ J $, we see that the two-particle continuum $ \Imag J_{pp} $ extends into the two-hole continuum and vice versa as shown in Fig.~\ref{J_autocoherent}. This is a general problem of the r-RPA approach.

\subsection{Contact}
A very interesting property of the occupation numbers is the
asymptotic behavior of the momentum distribution tails. In the case of
a contact interaction, the asymptotic behavior ($ k \gg
k_{\F}^{\sigma} $) follows the power law $ n^{\sigma}(k) \sim C/k^{4} $. The coefficient
\begin{equation}
  C = \lim_{k\to\infty} k^4 n^\uparrow(k) = \lim_{k\to\infty} k^4 n^\downarrow(k)
\end{equation}
is independent of the spin and, in the notation of Tan~\cite{Tan2008a,
  Tan2008b}, it is called \textit{contact}. This relationship has been
restated in a field theory context in
Ref.~\cite{Braaten2008}.\footnote{In Ref.~\cite{Braaten2008} a
  different notation is used where $C$ denotes Tan's contact
  integrated over the volume. Here we adopt Tan's notation, which is
  more convenient for uniform systems.} The value of the contact $ C
$ is related to different thermodynamic properties of the Fermi
gas~\cite{Tan2008b, Tan2008c}. For instance, it
  determines the dependence of the energy density $E/V$ on the
  interaction strength as \cite{Tan2008b}
  \begin{equation}
    \frac{d(E/V)}{d(-1/a)} = \frac{C}{4\pi m}\,. \label{eq:contact}
  \end{equation}

\begin{figure}[!h]
\includegraphics[scale=1.0]{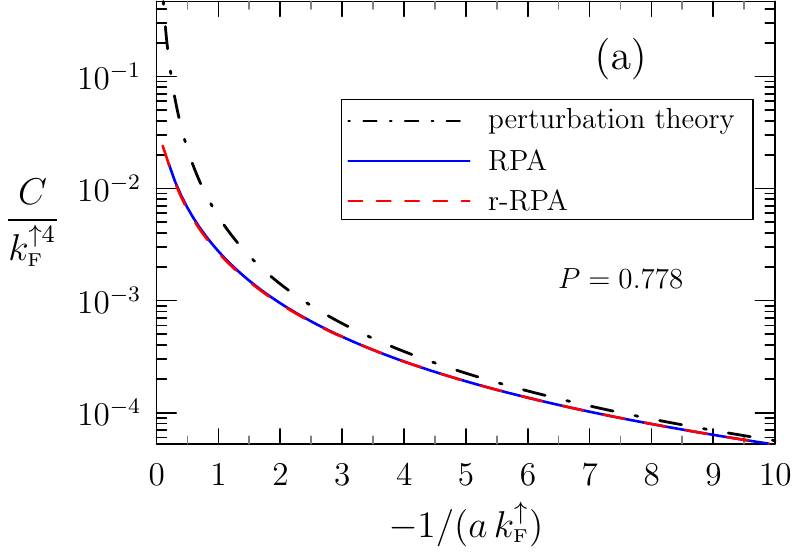} \\[0.2cm]
\includegraphics[scale=1.0]{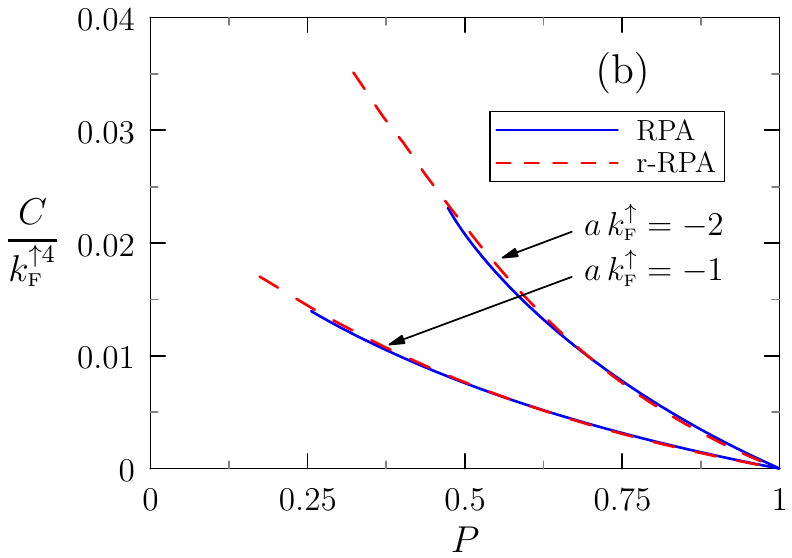}
\caption{Contact $ C $ (a) as a function of the interaction for an asymmetry of $ k_{\F}^{\downarrow} = k_{\F}^{\uparrow}/2 $; (b) as a function of the polarization for interactions of $ a \, k_{\F}^{\uparrow} = -1 $ and $ a \, k_{\F}^{\uparrow} = -2 $. The blue solid lines are the RPA results and the red dashed lines are for the r-RPA. The black dashed-dotted line represents the perturbative expression for the contact in the weakly interacting limit.}
\label{terme_contact_dependance}
\end{figure}

Figure~\ref{terme_contact_dependance} shows the dependence of the contact $ C $ on the interaction and polarization parameters for the RPA and the r-RPA.
Note that the value of the contact is almost identical within the RPA and the r-RPA, we do not know if the small difference between these curves is only due to the numerical precision or not. In the limit of weak interaction (large $ -1/(a \, k_{\F}^{\uparrow})) $, the contact approaches the perturbative result $ C = 16 \pi^{2} \, a^{2} \, \rho^{\uparrow} \rho^{\downarrow} $~\cite{Calvanese2018}, as can be seen in Fig.~\ref{terme_contact_dependance}.

\subsection{Occupation numbers calculated with the full Dyson equation}
The problem of the negative step that appears in the occupation numbers calculated with the standard RPA when the interaction becomes too strong is not improved by the self-consistent treatment. One way to cure this pathology is to use the complete Dyson equation instead of the truncated version~\eqref{fonction_green_ordre1} presented in \Sec{sec:Recall_of_pp-RPA}, i.e., to dress the Green function as
\begin{equation}\label{Equation_Dyson}
G = \dfrac{1}{1/G_{0} - \Sigma} \, .
\end{equation}
In terms of diagrams, the occupation numbers calculated with the complete Dyson equation are represented in Fig.~\ref{graphe_feynman}~(d).

\begin{figure}[!h]
\includegraphics[scale=0.8]{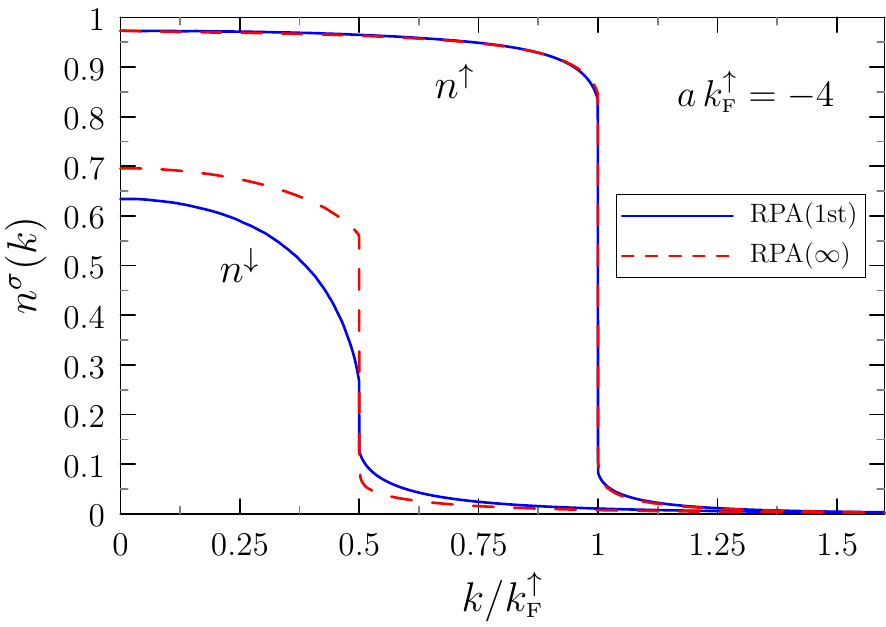}
\caption{The solid blue lines represent the occupation numbers calculated with RPA(1st), i.e., the truncated Dyson equation~\eqref{fonction_green_ordre1} and the dashed red lines represent the occupation numbers calculated with RPA($ \infty $), i.e., the complete Dyson equation~\eqref{Equation_Dyson}. The polarization is fixed at $ k_{\F}^{\downarrow} = k_{\F}^{\uparrow}/2 $, interaction strength is $ a \, k_{\F}^{\uparrow} = -4 $. The $ Z^{\downarrow} $ calculated with RPA($ \infty $) decreases much less than with the RPA(1st).}
\label{nombres_occupation_RPA_Dyson}
\end{figure}

When we consider the complete equation~\eqref{Equation_Dyson} to calculate the occupation numbers, the self-energy $ \Sigma $, Eq.~\eqref{eq:Sigma_general}, must be explicitly calculated, which is not the case when using the first-order truncated Dyson equation~\eqref{fonction_green_ordre1} as done before. The expressions for the imaginary part of the self-energy $ \Sigma = \Sigma_{hh} + \Sigma_{pp} $ are given by~\cite{Urban2014}
\begin{multline}\label{Sigma_hh}
\Imag \Sigma_{hh}^{\sigma} (\kk, \, \omega) = - \int_{p > k_{\F}^{\bar{\sigma}}} \dfrac{\mathrm{d}^{3} \mathbf{p}}{(2 \pi)^{3}} \, \theta(\Omega_{\F} - \omega - \varepsilon_{\pp}) \\
  \times \Imag \Gamma (\kk + \pp , \, \omega + \varepsilon_{\pp}) \, ,
\end{multline}
\begin{multline}\label{Sigma_pp}
\Imag \Sigma_{pp}^{\sigma} (\kk, \, \omega) = \int_{p < k_{\F}^{\bar{\sigma}}} \dfrac{\mathrm{d}^{3} \mathbf{p}}{(2 \pi)^{3}} \, \theta(\omega + \varepsilon_{\pp} - \Omega_{\F}) \\
  \times \Imag \Gamma (\kk + \pp , \, \omega + \varepsilon_{\pp})
\end{multline}
and the corresponding real parts are calculated with a dispersion relation. Then we calculate the spectral function
\begin{equation}\label{fonction_spectrale}
A^{\sigma} (\kk, \, \omega) =-\Imag  \dfrac{1}{\omega - \epsilon_{\kk} - \Sigma^{\sigma} (\kk, \, \omega - U^{\sigma})} \, .
\end{equation}
We have introduced the quantity $ U^{\sigma} = \Real \Sigma^{\sigma} (k_{\F}, \, \varepsilon_{\F}) $ to take into account the shift of the Fermi energy caused by the real part of $ \Sigma $. A useful property of the spectral function is
\begin{equation}\label{normalisation_fonction_spectrale}
\int_{\varepsilon_{\F} + U^{\sigma}}^{+ \infty} \dfrac{\mathrm{d}\omega}{\pi} \, A^{\sigma}(\kk, \, \omega) - \int_{- \infty}^{\varepsilon_{\F} + U^{\sigma}} \dfrac{\mathrm{d}\omega}{\pi} \, A^{\sigma}(\kk, \, \omega)  = 1 \, .
\end{equation}
The occupation numbers are obtained from
\begin{equation}
n^{\sigma} (k) = \pm \int_{- \infty}^{\varepsilon_{\F} + U^{\sigma}} \dfrac{\mathrm{d} \omega}{\pi} \, A^{\sigma} (\kk, \, \omega) \, ,
\end{equation}
with $ + $ for holes and $ - $ for particles. To avoid having to integrate the peak present in the spectral function when calculating the occupation numbers of the holes, we use Eq.~\eqref{normalisation_fonction_spectrale} that normalizes the spectral function and integrate over the complementary interval where there is no peak. Finally this leads to the formulas
\begin{equation}
n^{\sigma} (k < k_{\F}^{\sigma}) = 1 - \int_{\varepsilon_{\F} + U^{\sigma}}^{+ \infty} \dfrac{\mathrm{d} \omega}{\pi} \, A^{\sigma} (\kk, \, \omega) \, ,
\end{equation}
\begin{equation}
n^{\sigma} (k > k_{\F}^{\sigma}) = - \int_{- \infty}^{\varepsilon_{\F} + U^{\sigma}} \dfrac{\mathrm{d} \omega}{\pi} \, A^{\sigma} (\kk, \, \omega) \, .
\end{equation}
By comparing the results of this method, denoted RPA($\infty$), and those of the standard RPA (RPA(1st)), we see in Fig.~\ref{nombres_occupation_RPA_Dyson} that the occupation numbers within the RPA($\infty$) are less modified than those within RPA(1st). In particular, the negative step disappears even at the strongest interactions. Roughly speaking, since $ Z_{1\text{st}} \simeq 1 + \mathrm{d} \Sigma / \mathrm{d} \omega $ and $ Z_{\infty} \simeq 1/(1 - \mathrm{d} \Sigma / \mathrm{d} \omega) $ the step heights of the two methods are related by $ Z_{1\text{st}} \simeq 2 - 1/Z_{\infty} $. For weak interactions, both methods give similar results.

\begin{figure}[!h]
\includegraphics[scale=0.8]{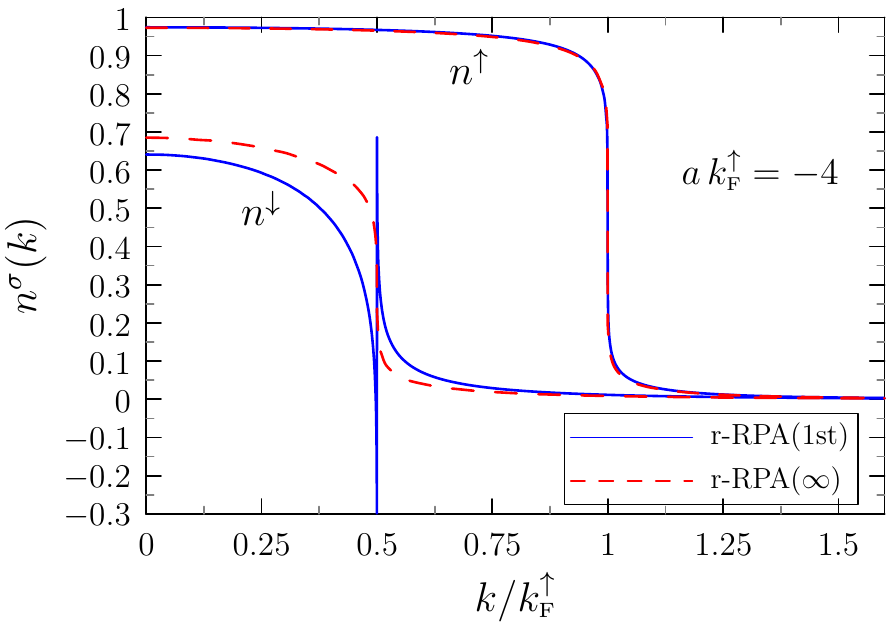}
\caption{Same as Fig.~\ref{nombres_occupation_RPA_Dyson} but using self-consistent occupation numbers (r-RPA). $ Z^{\downarrow} $ calculated with r-RPA($ \infty $) is positive whereas the one calculated with r-RPA(1st) is negative.}
\label{nombres_occupation_r-RPA_r-Dyson}
\end{figure}

To set up self-consistency, we adopt the same approach as in \Sec{sec:Renormalized_pp-RPA}. Figure~\ref{nombres_occupation_r-RPA_r-Dyson} shows the difference of the occupation numbers calculated with the r-RPA(1st) and r-RPA($ \infty $).
Again, in the occupation numbers calculated with the full Dyson equation, the problem of the negative step (and even negative occupation numbers) present in the r-RPA(1st) has disappeared. 

However there is a price to pay. If we define $ \rho_{\sigma}^{\Lut} = k_{\F}^{\sigma \, 3}/(6 \pi^{2}) $, the Luttinger theorem~\cite{Luttinger1960} states that the density $ \rho_{\sigma} $ calculated as the integral of the correlated occupation numbers satisfies the relation $ \rho_{\sigma} = \rho_{\sigma}^{\Lut} $. As shown in~\cite{Urban2014} this is exactly fulfilled within RPA(1st), but it is no longer true if the occupation numbers are calculated with the full Dyson equation (RPA($ \infty $)). More quantitatively, we define the relative error $ \Delta_{\text{\tiny{rel}}}^{\sigma} = |\rho_{\sigma}^{\Lut} -  \rho_{\sigma}|/ \rho_{\sigma} $. Table~\ref{tab_erreur} shows the violation of the Luttinger theorem for one specific example ($ a \, k_{\F}^{\uparrow} = -2.5 $, $ k_{\F}^{\downarrow} = k_{\F}^{\uparrow}/2 $).
\begin{table}
  \caption{\label{tab_erreur}Relative error $ \Delta_{\text{\tiny{rel}}}^{\sigma} $ (violation of the Luttinger theorem) for different calculations with $ a \, k_\F^{\uparrow} = -2.5 $ and $ k_{\F}^{\downarrow} = k_{\F}^{\uparrow}/2 $}
  \begin{ruledtabular}
    \begin{tabular}{ccccc}
               & RPA(1st) & r-RPA(1st) & RPA($ \infty $) & r-RPA($ \infty $)\\ \hline
      $ \Delta_{\text{\tiny{rel}}}^{\uparrow} $   & $ 0 $ & $ 0.053 \, \% $ & $ 0.073 \, \% $  & $ 0.013 \, \% $ \\
      $ \Delta_{\text{\tiny{rel}}}^{\downarrow} $ &  $ 0 $ & $ 0.45 \, \% $ & $ 2.9 \, \% $ & $ 2.1 \, \% $ \\ 
    \end{tabular}
  \end{ruledtabular}
\end{table}
We see that only the RPA(1st) satisfies the Luttinger theorem exactly. However, the r-RPA(1st) violates the Luttinger theorem only very slightly and the error observed can be due to the accumulation of numerical errors. The RPA($\infty$), on the contrary, clearly violates the Luttinger theorem and therefore the r-RPA($\infty$), too. At stronger interactions, the violation within RPA($\infty$) and r-RPA($\infty$) can be much worse, e.g., near the critical polarization at the unitary limit (see \Sec{subsec:phase_diagram}).

\section{Critical polarization}
\label{sec:Critical_polarization}
\subsection{FFLO transition}
\label{sec:FFLO}
\begin{figure}[!h]
\includegraphics[scale=0.8]{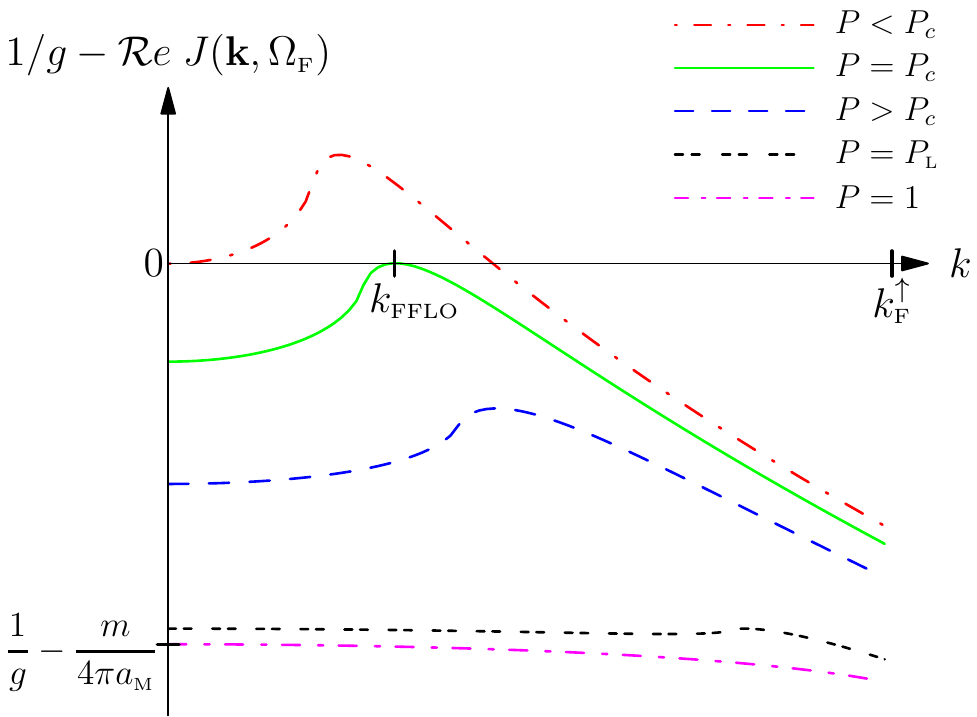}
\caption{Typical behavior of $ 1/g - J(\kk , \Omega_{\F})
    $ (computed within RPA for $ a \, k_{\F}^{\uparrow} = -1.913 $) as
  function of $ k $ for five polarizations $ P $. At the
  three highest polarizations associated with lower three (blue, black
  and purple) curves, the system is in the normal phase.  The green
solid line is at the critical polarization $P_c$ where the function
vanishes for a value of $ k $ which corresponds to $
k_{\FFLO} $. The red dash-dotted line is obtained for the
value of the polarization which gives the pole at $ \kk =
\boldsymbol{0} $ (BCS superfluidity), but it lies already in the FFLO
superfluid region. For the definitions of $P_{\Lif}$ and
$a_{\M}$, see text.}
\label{maximum_gJ}
\end{figure}

The Thouless criterion~\cite{Thouless1960} states that the superfluid transition occurs when a pole appears in the \textit{T} matrix at $ \omega = \Omega_{\F} $, i.e., $ 1/g - J (\kk, \omega = \Omega_{\F}) = 0 $. As long as the condition $ 1/g - J (\kk, \Omega_{\F}) < 0 $ is fulfilled for all $ \kk $, we are in the normal phase as shown in Fig.~\ref{maximum_gJ} as the blue dashed curve.

In the case of a non-polarized gas at finite temperature, approaching
$ T_{c} $ from above, the instability of the normal phase sets in
first at $ \kk = \boldsymbol{0} $. However, in the case of a polarized
gas, the difference between the Fermi levels favors the creation of
pairs with non-zero total momentum, resulting in the emergence of a
new type of superfluidity called FFLO
phase~\cite{Fulde1964,Larkin1964}. Now we look for the appearance of
the pole when approaching $ P_{c} $ from above. Considering $ 1/g - J
(\kk, \Omega_{\F}) $ as function of $ k $ for different values of $ P
$, as shown in Fig.~\ref{maximum_gJ}, we notice that the pole at $ \kk
\neq \boldsymbol{0} $ appears at a higher polarization than the one at
$ \kk = \boldsymbol{0} $. The value of $ k $ at which the pole appears
first, coming from strong polarization, is denoted $ k_{\FFLO} $:
\begin{equation}
1/g - \Real J (\mathbf{k}_{\FFLO}, \Omega_{\F}) = \max_{k} \left(1/g - \Real J (\kk, \Omega_{\F}) \right) \, .
\end{equation}
To determine the critical polarization, we must therefore determine the value of $ P $ such that 
\begin{equation}
1/g - \Real J (\mathbf{k}_{\FFLO}, \Omega_{\F}) = 0 \, .
\end{equation}

At the critical polarization, there is an instability of
  the system towards a formation of pairs with momentum
  $k_{\FFLO}$. Therefore $k_{\FFLO}$ corresponds more or less to the
  wave vector of the order-parameter oscillations in the FFLO
  phase. However, our theory does not tell us whether the paired phase
  that will be formed corresponds to a Fulde-Ferrel (FF) state with
  just one wave vector \cite{Larkin1964}, a Larkin-Ovchinnikov (LO)
  state with spatial modulations of the order parameter but without a
  varying phase \cite{Fulde1964}, or even more complicated states with
  a crystal-like structure.

\begin{figure}[!h]
\includegraphics[scale=1.0]{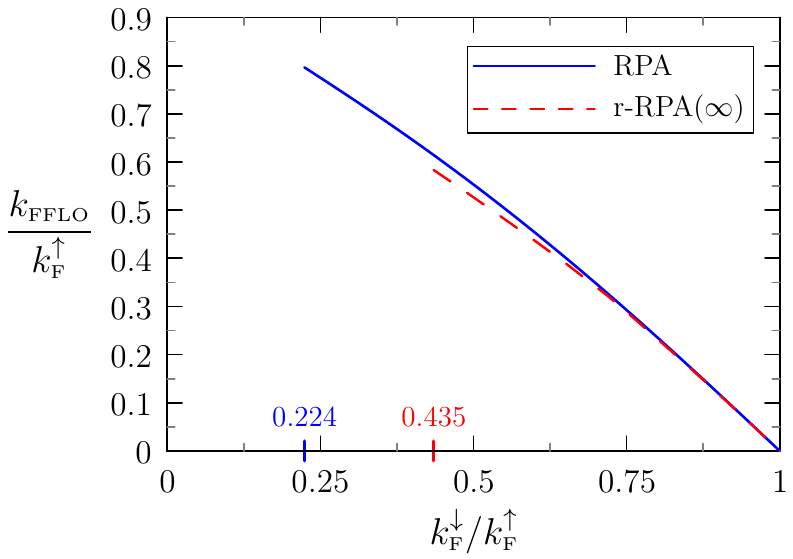}
\caption{Value of $ k_{\FFLO} $ as a function of $
  k_{\F}^{\downarrow}/k_{\F}^{\uparrow} $ within RPA (solid blue line)
  and within r-RPA($ \infty $) (dashed red line). In RPA, it is independent
  of the strength of the interaction. For the r-RPA($ \infty $), we have used $ a
  \, k_{\F}^{\uparrow} $ for each $ k_{\F}^{\downarrow}/k_{\F}^{\uparrow} $ such that $ P = P_{c} $. The FFLO phase disappears for $
  k_{\F}^{\downarrow}/k_{\F}^{\uparrow} < 0.224 $ in the RPA, and for $ k_{\F}^{\downarrow}/k_{\F}^{\uparrow} < 0.435 $ in the r-RPA($ \infty $).}
\label{graphe_k_fflo}
\end{figure}

  The qualitative picture described above and illustrated in
  Fig.~\ref{maximum_gJ} remains the same in a large range of
  interaction strengths $1/(a k_{\F})$. Increasing $1/(a k_{\F})$,
  i.e., $1/g$, simply shifts the curves upwards and thereby increases
  the critical polarization, which in turn leads to a higher value of
  $k_{\FFLO}$. The dependence of $k_{\FFLO}$ computed within RPA as a
  function of the ratio $ k_{\F}^{\downarrow}/k_{\F}^{\uparrow} $
  (which is directly related to $P$, small values of $
  k_{\F}^{\downarrow}/k_{\F}^{\uparrow} $ corresponding to large $P$
  and vice versa) is displayed in Fig.~\ref{graphe_k_fflo} as the
  solid blue line.

  At some positive value of $1/g$, corresponding to a strong critical
  polarization, the picture changes. As it can be seen from the
  double-dashed black curve in Fig.~\ref{maximum_gJ}, at some strong
  polarization denoted $P_{\Lif}$, the local maximum at $k=k_{\FFLO}$
  has become so flat that its value coincides with another local
  maximum that has built up at $k=0$. Hence, at $P=P_{\Lif}$, the global
  maximum of $1/g-J$ changes from the one at $k\neq 0$ to the one at
  $k=0$, and as a consequence, $k_{\FFLO}$ jumps from a finite value
  to zero, as can be seen in Fig.~\ref{graphe_k_fflo}. This
  corresponds to the Lifshitz point L in the schematic phase diagram
  of \cite{Son2006}.
    
\subsection{Implementation of self-consistency}
\begin{figure}[!h]
\includegraphics[scale=0.8]{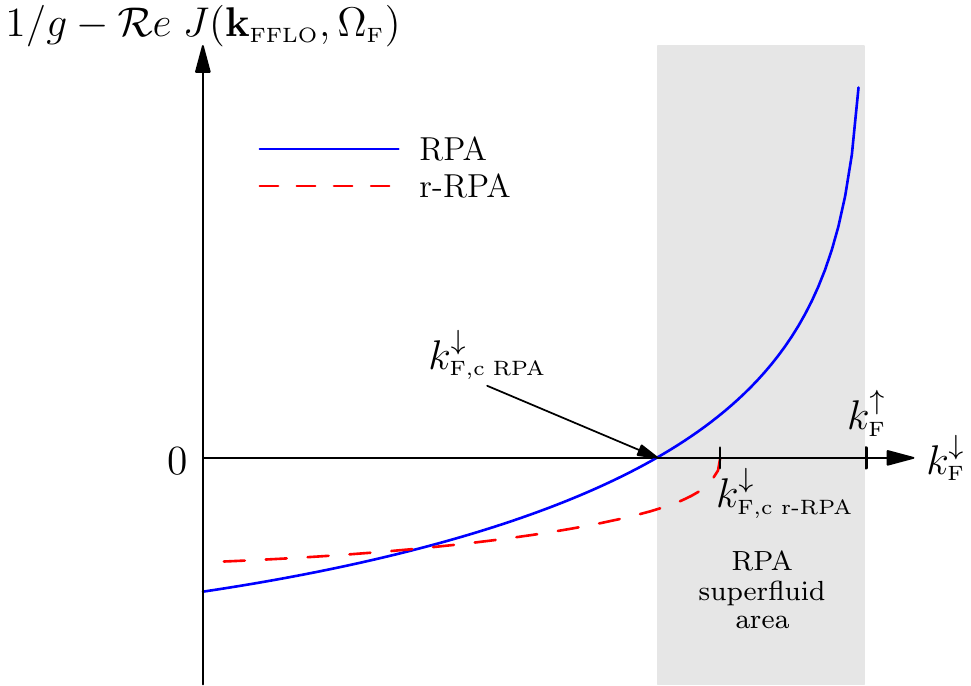}
\caption{Schematic behavior of the maximum value of the
    function $ 1/g - \Real J (\kk, \Omega_{\F}) $ depending on $
    k_{\F}^{\downarrow} $ for fixed $ k_{\F}^{\uparrow} $. The blue
    solid line is obtained from the RPA and the red dashed line from
    the r-RPA. As can be seen, the critical polarization obtained from
    r-RPA is in the superfluid phase of the RPA.}
\label{schema_convergence}
\end{figure}

As explained in the preceding subsection, for fixed
  values of $ k_{\F}^{\uparrow} $ and $a$, the method to find the
critical polarization is to determine the zero of $ 1/g - J
(\kk_{\FFLO}, \Omega_{\F}) $ as a function of $ P $, i.e., in practice
as a function of $ k_{\F}^{\downarrow} $. This curve is shown in
Fig.~\ref{schema_convergence}.
The value of $ k_{\F}^{\downarrow} $ corresponding to $ P_{c} $ will
be denoted $ k_{\Fc}^{\downarrow} $. For values of $
k_{\F}^{\downarrow}$ lower than $ k_{\Fc}^{\downarrow}
$, we are in the normal phase, otherwise we are in the superfluid
phase which cannot be described with our theory.

\begin{figure}[!h]
\includegraphics[scale=1.0]{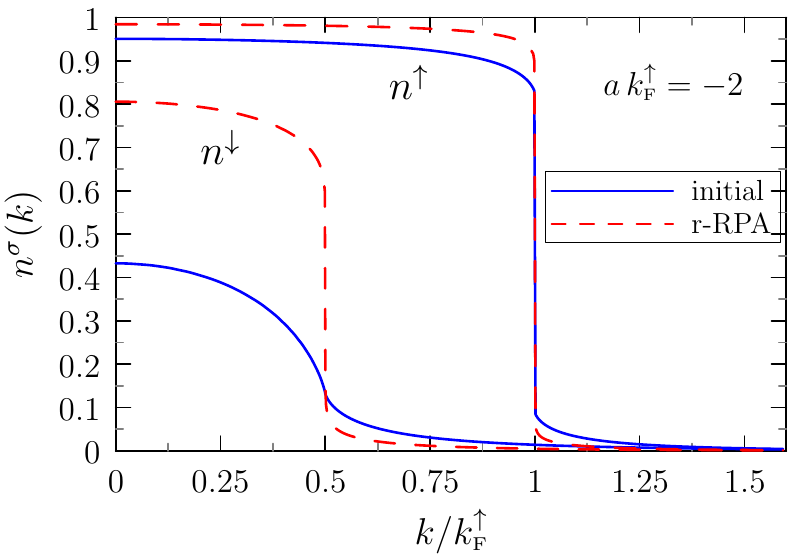}
\caption{Up and down occupation numbers for the polarization $ k_{\F}^{\downarrow} = k_{\F}^{\uparrow}/2 $ ($ P = 0.778 $) and $ a \, k_{\F}^{\uparrow} = -2 $. Solid blue lines represent the occupation numbers used for the initialization of the self-consistent iteration. Dashed red lines represent the converged occupation numbers within the r-RPA(1st) calculation.}
\label{nombres_occupation_modifies}
\end{figure}

However, it is impossible to compute this curve for the r-RPA up to
the corresponding $ k_{\Fc}^{\downarrow} $ (for given $ a $ and $
k_{\F}^{\uparrow} $) if one initializes the self-consistent iteration
with the uncorrelated occupation numbers (step functions). The reason
becomes clear from Fig.~\ref{schema_convergence}: with increasing
correlations, $ P_{c} $ becomes smaller, i.e., $ k_{\Fc}^{\downarrow}
$ becomes larger. Therefore, $ k_{\Fc}^{\downarrow} $ for r-RPA lies
in the superfluid area of RPA and already the first iteration step
cannot be performed.  To avoid going through the superfluid zone, we
need to find initial conditions that maximize the correlations and at
the same time $ k_{\text{\tiny F,c}}^{\downarrow} $ so that it
decreases with each iteration. Fortunately the result is independent
of the initial occupation numbers. The occupation numbers for the
initialization are constructed from those calculated with the RPA by
artificially increasing as much as possible the correlated part $
\delta n^{\sigma}(k) = n^{\sigma}(k) - \theta(k_{\F}^{\sigma} - k) $
such that $ Z^{\downarrow} = 0 $
(cf. Fig.~\ref{nombres_occupation_modifies}).

We then use the procedure described in the preceding section to find the $ k_{\FFLO} $. In the r-RPA($ \infty $), we observe a more and more important deviation from the RPA as the polarization increases as can be seen by comparing the solid blue and the dashed red lines in~Fig.~\ref{graphe_k_fflo}. Also the critical value of $ k_{\F}^{\downarrow} $ changes (where $ k_{\FFLO} $ disappears). In RPA, we
find this point at $ (k_{\F}^{\downarrow}/k_{\F}^{\uparrow})_{\Lif} = 0.224 $ and slightly on the BEC side
close to unitarity at $ 1/(a_{\Lif} k_{\F}^{\uparrow}) \simeq 0.210 $,
in agreement with the disappearence of the FFLO phase in mean-field
theory, see Fig.~6(a) of~\cite{Calvanese2018}. In r-RPA($ \infty $), we find $ (k_{\F}^{\downarrow}/k_{\F}^{\uparrow})_{\Lif} = 0.435 $ at $ 1/(a_{\Lif} k_{\F}^{\uparrow}) \simeq 0.131 $ which is closer to the unitarity compared to the RPA result. It is due to the fact that the maximum that gives the $ k_{\FFLO} $ becomes flatter with the self-consistency. Therefore, it drops more quickly below the global maximum at $ k = 0 $ and the FFLO phase is lost.

\subsection{Phase diagram}
\label{subsec:phase_diagram}
\begin{figure}[!h]
\includegraphics[scale=0.85]{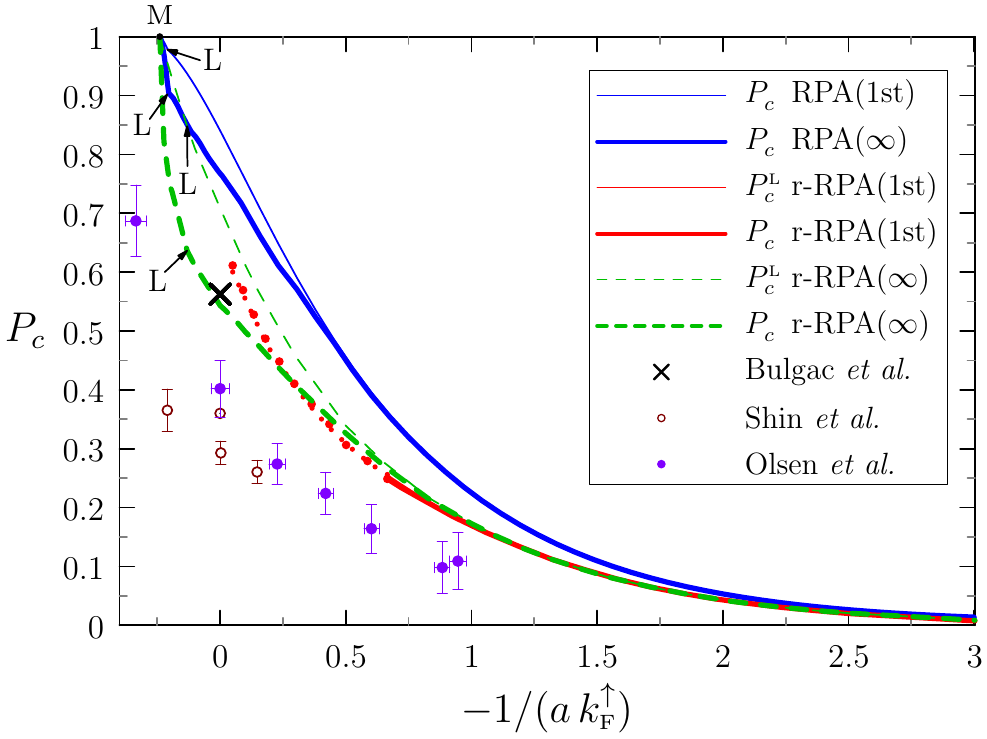}
\caption{Critical polarization from the Thouless
    criterion as function of $ -1/(a \, k_{\F}^{\uparrow}) $. Thick
  lines correspond to $ P_{c} $ calculated with the occupation numbers
  and thin lines to $ P_{c}^{\Lut} $ that one would obtain if the
  Luttinger theorem was satisfied. The blue solid lines represent the
  RPA calculation. The red lines are for r-RPA(1st), in the range
  where they are dotted it gives a negative step ($ Z < 0 $). The
  green dashed lines are for r-RPA($ \infty $). The black
    cross is the theoretical prediction extracted from Fig.~1 of
    Ref.~\cite{Bulgac2008}. We also show experimental results for the
    limit of the coexistence region corresponding to a first-order
    phase transition towards an unpolarized superfluid phase; the data
    are from Refs.~\cite{Shin2008,Shin2008prl} (Shin et al.) and
    \cite{Olsen2015} (Olsen et al.).}
\label{diagramme_phase_convergence_Dyson}
\end{figure}

Following the procedure discussed in the preceding
  subsections, we are only able to detect the instability of the
  normal phase corresponding to a second-order phase transition to the
  superfluid phase. However, if there is a first-order phase
  transition from the normal to the superfluid phase with a
  coexistence region extending to a higher polarization than our
  critical polarization $P_c$, we are not able to see it, because this
  would require to compute the energy of the superfluid phase. Keeping
  this word of caution in mind, we show in
  Fig.~\ref{diagramme_phase_convergence_Dyson} the phase diagram
  giving the critical polarization as a function of the interaction
  strength.

  Notice that on the BEC side of the crossover, beyond
    some interaction $-1/(a_{\M} k_{\F}^{\uparrow}) < 0$, the
    polarized normal phase (except at $P=1$ where the system is
    non-interacting) does not exist any more. This point, denoted M,
    corresponds to the polaron-to-molecule transition discussed in
    Refs.~\cite{Prokofev2008,Punk2009} where few down particles added
    to a fully polarized system of up particles do not form a normal
    fluid Fermi sea of polarons any more, but a BEC of molecules with
    total momentum $k=0$. In the present framework, this transition
    happens when the lowest curve shown in Fig.~\ref{maximum_gJ},
    corresponding to $P=1$, is shifted upwards to the horizontal axis,
    i.e., at
  \begin{multline}
    \frac{1}{a_{\M}k_{\F}^{\uparrow}}
    = \frac{4\pi}{mk_{\F}^{\uparrow}}J(0,\varepsilon_{\F};k_{\F}^{\downarrow}=0)
    = \frac{2}{\pi}-\frac{1}{\sqrt{2}\pi}\ln\frac{\sqrt{2}+1}{\sqrt{2}-1}
    \\ \simeq 0.24\,.
  \end{multline}
  This value has to be compared with the exact one,
  $1/(a_{\M}k_{\F}^{\uparrow})\simeq 0.9$ \cite{Prokofev2008}. The origin of
  this discrepancy is that the ladder approximation includes only
  2-particle (2p) but no 3-particle-1-hole (3p1h) and more complicated
  states \cite{Punk2009}. Since at $P_c=1$ no down particles are
  present any more, the self-consistency does not have any effect and
  all our curves, whether they are calculated with RPA or r-RPA, end
  in the same point M.

  Next to the point M, we have the transition towards the $k=0$
  superfluid phase. Only to the right of the Lifshitz point L mentioned in
  Sec.~\ref{sec:FFLO}, visible as kink in the curves pointed by the black arrows in
  Fig.~\ref{diagramme_phase_convergence_Dyson}, the transition is of
  the FFLO type.

We see that, as expected, the critical polarization in r-RPA is always
lower than in RPA. Notice that the r-RPA(1st) (red lines) breaks down
at interactions stronger than $ 1/(a \, \smash{k_{\F}^{\uparrow}})
\simeq -0.675 $ where a negative step appears in the occupation
numbers. Nevertheless, it is still possible to reach convergence in some range and these results are shown as red dotted
lines in Fig.~\ref{diagramme_phase_convergence_Dyson}. In contrast to
the r-RPA(1st), the r-RPA($ \infty $), shown as the green dashed
lines, allows us to describe the whole range of interactions up to the
point M, since a negative step does not appear. At weak
interaction, the computation with r-RPA($ \infty $) gives the same
result as r-RPA(1st) which is obvious since the occupation numbers are
practically the same in this case.

The critical polarization within RPA is actually the same
  as within mean field including the possibility of the FFLO phase but
  no phase separation~\cite{Pao2009} (see also Fig.~6(a)
  of~\cite{Calvanese2018}).\footnote{In~\cite{Pao2009} and in Fig.~6(a)
    of~\cite{Calvanese2018}, $P_c$ is shown as a function of
    $1/(a \, k_{\F}) = (1+P)^{1/3}/(a \, k_{\F}^\uparrow)$.} In
  particular, in the unitary limit, the RPA predicts $ P_{c} = 0.834
  $. This is clearly higher than the critical polarization $ P_{c} =
  0.562$ obtained from an energy-density functional (ASLDA) fitted to
  Quantum-Monte-Carlo (QMC) results~\cite{Bulgac2008}, marked by the
  cross in Fig.~\ref{diagramme_phase_convergence_Dyson}. Thus the
r-RPA($ \infty $) result of $ P_{c} = 0.543 $ (thick dashed line)
would be a significant improvement compared to the RPA. However this
low $ P_{c} $ is to some extent a consequence of the violation of the
Luttinger theorem. In fact, if one computes $ P_{c} $ under the
assumption that the Luttinger theorem is satisfied, i.e., $
P_{c}^{\Lut} = (k_{\F}^{\uparrow 3} - k_{\F}^{\downarrow
  3})/(k_{\F}^{\uparrow 3} + k_{\F}^{\downarrow 3}) $, one obtains
only a weaker reduction from $ 0.834 $ to $ 0.709 $ (thin dashed
line).

Since in RPA(1st) the Luttinger theorem is exactly fulfilled, we have
$ P_{c}^{\Lut}(\text{RPA(1st)}) = P_{c}(\text{RPA(1st)}) $ (thin blue
line). Since the RPA($ \infty $) uses the same uncorrelated occupation
numbers as the RPA(1st), it is clear that also $
P_{c}^{\Lut}(\text{RPA($\infty$)}) = P_{c}(\text{RPA(1st)}) $. The
reduction of $ P_{c}(\text{RPA($\infty$)}) $ (thick blue line) near
unitarity is therefore only due to the violation of the Luttinger
theorem. The violation of the Luttinger theorem in RPA($ \infty $) and
r-RPA($ \infty $) gets stronger and stronger as we approach the
unitary limit. It becomes visible in the region of interaction and
polarization parameters where the RPA(1st) and r-RPA(1st) give a
negative step. On the contrary, for the r-RPA(1st), we have $
P_{c}(\text{r-RPA(1st)}) \simeq P_{c}^{\Lut}(\text{r-RPA(1st)}) $ for
all values of interaction, therefore we tend to suspect that the
Luttinger theorem is fulfilled for the r-RPA(1st) and that the small
discrepancy comes from the accumulation of numerical errors. Finally,
treating the RPA with the Dyson equation avoids occupation numbers
with negative steps but there are still unphysical features.

Finally, let us say a few words about the phase separation
  observed in \cite{Shin2008,Shin2008prl,Olsen2015}. These experiments
  indicate that at the polarizations shown by the circles in
  Fig.~\ref{diagramme_phase_convergence_Dyson}, there is a first-order
  phase transition towards an unpolarized superfluid, in good
  agreement with QMC results which did not include the possibility of
  FFLO-type phases \cite{Pilati2008}. The experimental critical
  polarization for phase separation is much lower than the one
  obtained in mean field \cite{Pao2009} and leaves some room for a
  possible continuous (second order) transition towards the FFLO phase
  before phase separation happens, which corresponds qualitatively to
  the scenario suggested in \cite{Bulgac2008} for the case of
  unitarity.

\section{Conclusions}
\label{sec:conclusions}
We have implemented a self-consistent calculation of the occupation
numbers in the formalism of the r-RPA. This is an
  approximation to the so-called self-consistent RPA (by RPA we mean
  the RPA in the particle-particle channel, pp-RPA). On the one hand,
in the context of polarized Fermi gases, the r-RPA is
interesting because it reduces the critical polarization at $ T = 0 $
that is strongly overestimated with RPA. The r-RPA is therefore an
improvement in the description of this system. On the other hand, the
formalism does not cure certain pathologies specific to
the RPA. One of the problems of standard RPA [RPA(1st)]
  is that in the strongly coupled regime the quasiparticle $Z$ factor
  at the Fermi surface (i.e., the height of the step in the occupation
  numbers) becomes negative. A way to cure this pathology is to
consider the complete Dyson equation [RPA($\infty$)]
  instead of the truncated one which is used in RPA(1st) (and also
  in the NSR approach at finite temperature). This method does not
conserve the number of particles (Luttinger theorem) but
it gives at least physical occupation numbers for arbitrary strength
of interaction.

Other self-consistent calculations exist to treat in-medium
correlations in cold atoms. These numerically demanding methods are
based on the self-consistent Green's functions also called
Luttinger-Ward formalism at finite temperature. These methods were
applied to study the finite temperature BCS-BEC
crossover, in particular the phase diagram of
non-polarized~\cite{Haussmann1994, Haussmann2007} and
polarized~\cite{Frank2018} gases. In~\cite{Pieri2017}, it was shown
that within the Luttinger-Ward formalism, the Luttinger theorem is
exactly fulfilled with the full Dyson equation.

The r-RPA method allows us to see the evolution of 
occupation numbers by including correlations in the ground state. 
An interesting aspect is that the tails of the occupation numbers, 
whose information is contained in the contact term, remains almost 
unchanged with the inclusion of self-consistency.

We discussed the transition to the superfluid phase as an
  instability in the $\textit{T}$ matrix (Thouless criterion). Except
  at very strong polarizations (beyond the Lifshitz point), the
  transition is predicted to be of the FFLO type and our r-RPA result
  for the corresponding critical polarization in the unitary limit is
  close to the one of \Ref{Bulgac2008}.

  Although the FFLO phase has attracted a lot of attention (also in
the context of imbalanced gases with two components of different
masses~\cite{Forbes2005}), its existence is still an open
  question. In particular, there is a competition between phase
  separation (as seen in experiments) and the formation of the FFLO
  phase~\cite{Bulgac2008, Bedaque2003, Liu2003, Son2006}. With the
  present theory, which is limited to the normal phase, we cannot
  address the question of phase separation because it requires the
  calculation of the energy in the superfluid phase. The FFLO phase
  and phase separation can perhaps be reconciled by interpreting the
  FFLO phase (to be precise, the LO phase) as a periodic ``micro-phase
  separation''~\cite{Radzihovsky2010}. Furthermore, as it was pointed
  out, e.g., in~\cite{Radzihovsky2010}, the harmonic potential of the
  trap makes it difficult to see the FFLO phase in
  experiments. Hopefully, future experiments with flat traps as they
  are being built will clarify this question.
\section*{Acknowledgments}
\label{sec:acknowledgments}

We thank Peter Schuck for useful discussions and for
  comments on the manuscript.

\bibliography{biblio}

\end{document}